\documentclass[lettersize,journal]{IEEEtran}
\usepackage{amsmath,amsfonts}
\usepackage{algorithmic}
\usepackage{algorithm}
\usepackage{array}
\usepackage[caption=false,font=normalsize,labelfont=sf,textfont=sf]{subfig}
\usepackage{textcomp}
\usepackage{stfloats}
\usepackage{url}
\usepackage{verbatim}
\usepackage{graphicx}
\usepackage{cite}
\usepackage{orcidlink}
\usepackage{multirow}
\begin{document}

\title{\textbf{SPADE}: A SIMD Posit-enabled compute engine\\ for Accelerating DNN Efficiency}

\author{Sonu Kumar\orcidlink{0009-0000-4008-7153}, 
Lavanya Vinnakota\orcidlink{}, Mukul Lokhande \orcidlink{0009-0001-8903-5159}, \IEEEmembership{Member, IEEE}\\ Santosh Kumar Vishvakarma\textsuperscript{$\ast$}\orcidlink{0000-0003-4223-0077}, \IEEEmembership{Senior Member, IEEE}, and Adam Teman\orcidlink{0000-0002-8233-4711}, \IEEEmembership{Senior Member, IEEE}, 

\thanks{Sonu Kumar and Santosh Kumar Vishvakarma thank DST for the INSPIRE PhD fellowship, and MeitY/SMDP-C2S for ASIC tools support. Sonu Kumar is with the Centre for Advanced Electronics, IIT Indore. Lavanya Vinnakota and Mukul Lokhande are with the NSDCS Research Group, IIT Indore. \\\textbf{Corr. author\textsuperscript{$\ast$}}: Santosh Kumar Vishvakarma (skvishvakarma@iiti.ac.in).}
\thanks{Manuscript received Mon XX, 202x; revised Mon XX, 202x.}}

\markboth{IEEE Transactions on Very Large Scale Integration Systems,~Vol.~XX, No.~X, Month~202x}%
{Kumar \MakeLowercase{\textit{et al.}}: A SIMD Posit-enabled compute engine for Accelerating DNN Efficiency}


\maketitle

\begin{abstract}
The growing demand for edge-AI systems requires arithmetic units that balance numerical precision, energy efficiency, and compact hardware while supporting diverse formats. Posit arithmetic offers advantages over floating- and fixed-point representations through its tapered precision, wide dynamic range, and improved numerical robustness. This work presents SPADE, a unified multi-precision SIMD Posit-based multiply-accumulate (MAC) architecture supporting Posit (8,0), Posit (16,1), and Posit (32,2) within a single framework. Unlike prior single-precision or floating/fixed-point SIMD MACs, SPADE introduces a regime-aware, lane-fused SIMD Posit datapath that hierarchically reuses Posit-specific submodules (LOD, complementor, shifter, and multiplier) across 8/16/32-bit precisions without datapath replication. FPGA implementation on a Xilinx Virtex-7 shows 45.13\% LUT and 80\% slice reduction for Posit (8,0), and up to 28.44\% and 17.47\% improvement for Posit (16,1) and Posit (32,2) over prior work, with only 6.9\% LUT and 14.9\% register overhead for multi-precision support. ASIC results across TSMC nodes achieve 1.38 GHz at 6.1 mW (28 nm). Evaluation on MNIST, CIFAR-10/100, and alphabet datasets confirms competitive inference accuracy.
\end{abstract}

\begin{IEEEkeywords}
Posit MAC, deep learning accelerators, multi-precision systolic arrays, single instruction, multiple data (SIMD) processing elements.
\end{IEEEkeywords}

\section{Introduction}
\IEEEPARstart{T}{he} rapid proliferation of edge-AI systems and deep neural network (DNN) workloads has intensified the demand for arithmetic units that simultaneously offer high numerical accuracy, low power consumption, and compact hardware realization. Among compute kernels, the Multiply-Accumulate (MAC) operation dominates both execution time and energy consumption, making its efficiency a critical design concern for next-generation accelerators \cite{QuantMAC}. As DNNs continue to scale in depth and complexity, practical edge deployment increasingly requires an explicit balance among numerical precision, throughput, and silicon cost.

Conventional number formats provide only partial solutions under tight area and energy constraints. Fixed-point arithmetic is area- and power-efficient, but its limited dynamic range and scale selection can introduce numerical instability and accuracy loss \cite{Flex-PE}. Floating-point arithmetic improves robustness through a wide dynamic range, yet incurs substantial hardware overhead from exponent alignment, normalization, rounding, and special-case support, which often makes it expensive for always-on edge devices \cite{RAMAN_VLSID26, HYDRA_ICIIS25}.

Posit arithmetic has emerged as a promising alternative by providing tapered precision and a wide dynamic range with compact encodings. Its self-scaling regime and flexible bit allocation can deliver high numerical fidelity at reduced bit widths compared to IEEE floating-point \cite{RPE}. Consequently, recent works have proposed single-precision Posit MAC datapaths optimized for fixed formats like Posit-8, Posit-16, or Posit-32 \cite{PDPU}. However, modern DNNs exhibit strong layer-wise precision heterogeneity, where some layers tolerate aggressive quantization while others demand higher fidelity, motivating multi-precision execution with single accelerator data-path \cite{Flex-PE}.

Despite this motivation, multi-precision SIMD MAC architectures have largely focused on fixed-point, floating-point, or bfloat formats, while the attractive numerical properties of Posit remain under-explored in SIMD multi-precision hardware. The main challenge is architectural: Posit computation requires variable-length regime decoding, precision-dependent alignment and normalization, and carefully managed carry propagation, all of which complicate efficient SIMD lane fusion and hardware sharing.

This work proposes \textit{SPADE} (SIMD Posit Compute Unit for Accelerated DNN Efficiency), a Posit-enabled multi-precision SIMD MAC architecture designed to support precision-adaptive DNN execution on resource-constrained platforms. SPADE introduces a hierarchical SIMD lane fusion strategy that enables independent Posit-8 lanes, paired Posit-16 lanes, and a unified Posit-32 datapath within a single compute engine. By reusing Posit-specific submodules across precisions, SPADE provides precision scalability with minimal control and area overhead, while preserving Posit’s numerical advantages for edge-AI workloads.

The primary contributions of this paper are:
\begin{itemize}
    \item \textbf{Hierarchical multi-precision SIMD Posit MAC architecture:} We propose a regime-aware SIMD lane fusion and sharing strategy that unifies Posit-8/16/32 execution within one datapath, enabling precision-adaptive MAC computation while reusing Posit-specific submodules.
    \item \textbf{Posit-specific design insights for efficient SIMD fusion:} We identify and address key implementation challenges unique to Posit arithmetic (regime handling, normalization, and precision-dependent carry/shift behavior) to enable scalable SIMD operation with low overhead.
    \item \textbf{End-to-end validation across hardware and workloads:} We validate SPADE using RTL-level implementation and FPGA prototyping, and we report synthesis and place-and-route results across multiple technology nodes, together with DNN-level evaluation to demonstrate the accuracy and efficiency tradeoffs enabled by precision adaptivity.
\end{itemize}

\begin{figure*}[!t]
    \centering
    \includegraphics[width=\textwidth]{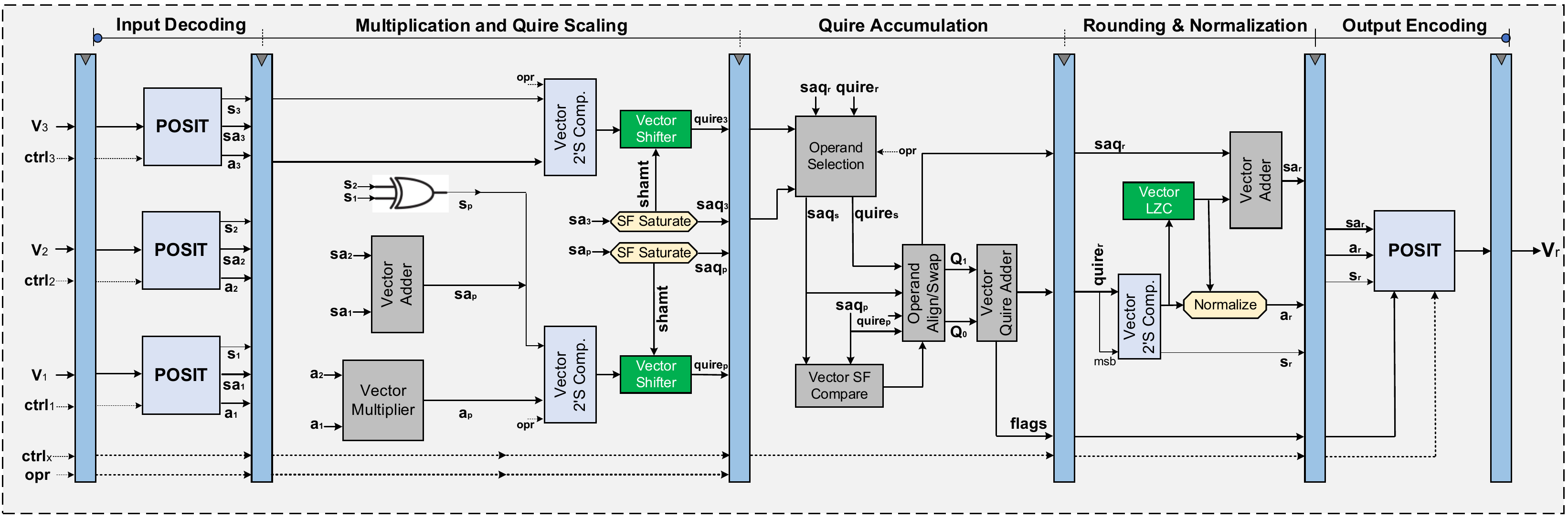}
    \caption{Proposed regime-aware SIMD Posit-8/16/32 MAC datapath illustrating hierarchical lane fusion and shared Posit-specific submodules.}
    \label{fig:posit-mac}
\end{figure*}

\section{Proposed Work}

\subsection{Background \& Motivation}

Deep neural network (DNN) workloads exhibit pronounced layer-wise precision heterogeneity. Early convolution layers are typically error-resilient and can tolerate aggressive quantisation, while deeper convolutional or fully connected layers demand higher numerical fidelity to preserve inference accuracy. Fixed-precision MAC architectures fail to effectively exploit this property, either overprovisioning hardware resources for low-precision layers or degrading accuracy when constrained to uniformly reduced precision. In the context of Posit arithmetic, standalone high-precision Posit-32 MAC units further exacerbate this inefficiency by exhibiting poor utilisation and excessive energy consumption when executing low-bitwidth workloads.

To address such inefficiencies, recent Neural Processing Units (NPUs) are increasingly adopting multi-precision datapaths that dynamically trade numerical precision for improved performance and energy efficiency. Commercial and academic accelerators exemplify this trend. For instance, the Samsung NPU integrated in a 4\,nm mobile SoC supports unified 8/16/32/64-bit floating-point execution modes \cite{Samsung-NPU}, while AMD’s XDNA NPU enables adaptive computation using FP, BF16, and TF32 formats \cite{XDNA}. Similarly, RISC-V-based accelerators such as Maestro \cite{Maestro} and Occamy \cite{Occamy} demonstrate scalable vector-tensor processing across 8- to 64-bit precision ranges, targeting both edge and high-performance computing workloads. Mini-float-based SIMD MAC architectures optimized for Versal MPSoCs have also been reported in \cite{MACC}.

While these designs clearly establish the importance of multi-precision execution, they predominantly target floating-point or bfloat formats. As a result, they often incur substantial area and power overheads due to LUT-intensive control logic, wide Booth-based multipliers, and complex normalization pipelines \cite{XRNPE_VLSID26}. Moreover, the architectural implications of Posit arithmetic such as variable-length regime decoding, precision-dependent alignment, and normalization remain largely unexplored in the context of multi-precision SIMD MAC architectures. These limitations motivate the need for a Posit-aware multi-precision SIMD MAC architecture that can exploit layer-wise precision heterogeneity without incurring the overheads associated with conventional floating-point designs. Such an architecture must efficiently support Posit-specific operations while enabling scalable precision adaptation within a unified datapath.

\begin{figure*}[!t]
    \centering
    \includegraphics[width=\textwidth]{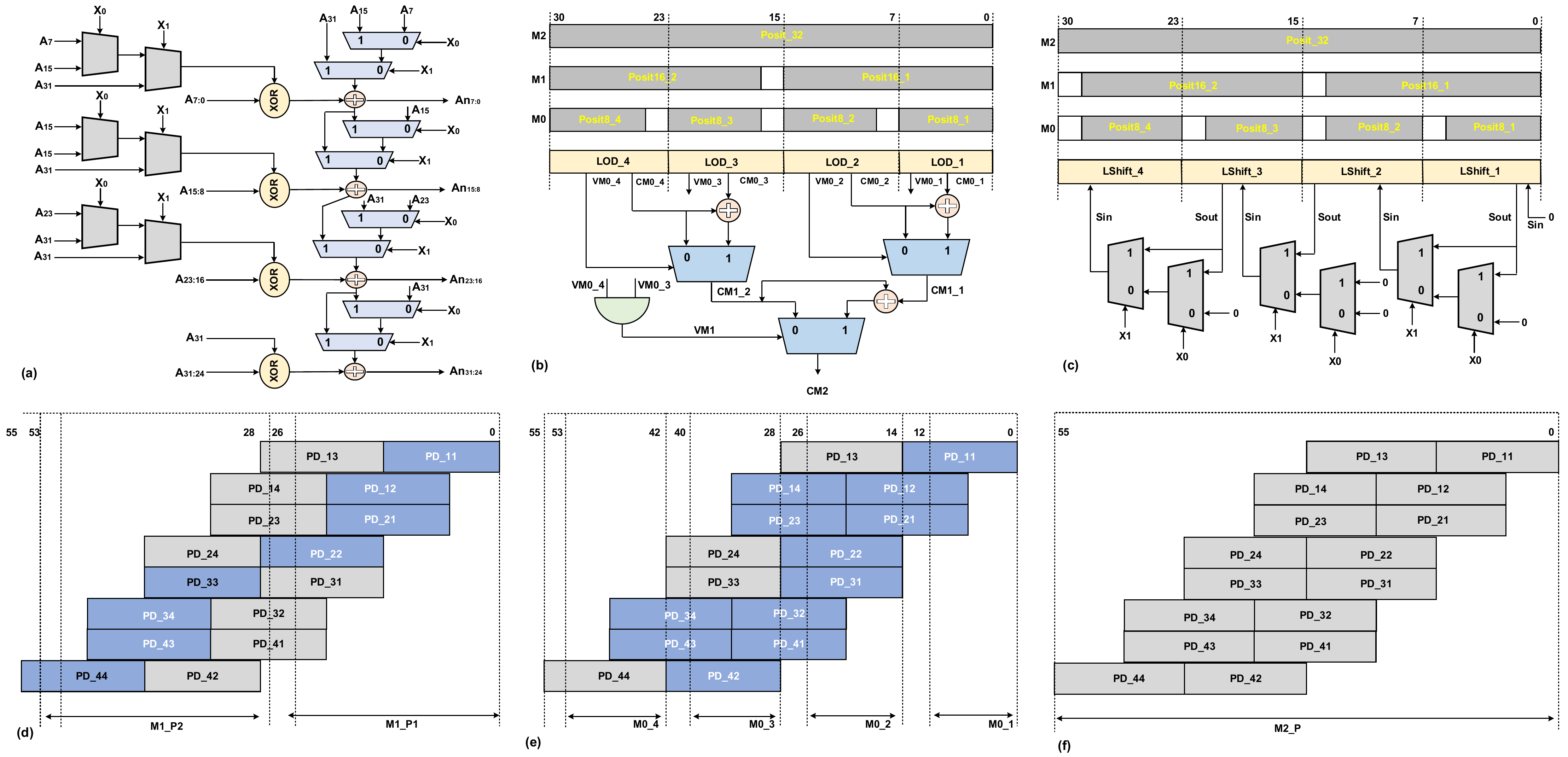}
    \caption{Detailed internal circuitry showcasing SIMD configurable circuit blocks: (a) Leading-One Detector, (b) complementor in the accumulation stage, (c) multi-stage logarithmic barrel shifter, and multiplier in (d) 8-bit, (e) 16-bit and (f) 32-bit partition mode.}
    \label{fig:comp_circ}
\end{figure*}


\subsection{SPADE MAC Engine}

The SPADE MAC engine follows a five-stage Posit MAC pipeline designed to support error-free accumulation and scalable SIMD execution across multiple Posit formats. The overall datapath is shown in Fig.~\ref{fig:posit-mac}.

\textbf{Stage 1: Posit Unpacking and Field Extraction}  
Each Posit operand is unpacked to extract the sign, regime, exponent, and mantissa fields. The sign bit determines whether two’s complementation is required. A Leading-One Detector (LOD) decodes the variable-length regime, after which the operand is left-shifted to extract the exponent and mantissa with an implicit leading one. The combined regime and exponent information is used to compute a scaling factor forwarded to the accumulation stage.

\textbf{Stage 2: Mantissa Multiplication}  
The extracted mantissas are multiplied using a precision-scalable SIMD multiplier. Although mantissa widths vary across Posit-8, Posit-16, and Posit-32 formats, the implicit leading bit ensures high-precision products. The resulting mantissa product is forwarded to the accumulation stage without intermediate rounding.

\textbf{Stage 3: Quire-Based Accumulation}  
The mantissa product is accumulated in a wide quire register, enabling exact accumulation without intermediate rounding. Products are aligned using scaling-factor differences before accumulation, while arithmetic right shifts preserve sign correctness. Accumulation is gated by an enable signal to support bypass when required, preserving the wide dynamic range of Posit arithmetic.

\textbf{Stage 4: Reconstruction and Normalization}  
The accumulated fixed-point value in the quire is converted back into Posit form through normalization using the SIMD LOD, followed by recomputation of the regime and exponent fields and extraction of the mantissa. This process ensures compatibility with the target Posit precision while maintaining numerical correctness.

\textbf{Stage 5: Rounding and Packing}  
The reconstructed mantissa is rounded using a round-to-nearest-even scheme based on guard, round, and sticky bits to minimize quantization error. The rounded fields are then packed into the final Posit word corresponding to the selected precision.

SPADE employs four precision-scalable SIMD submodules to enable multi-precision execution: the SIMD Complementor, SIMD LOD, SIMD Shifter, and SIMD Multiplier. A unified 2-bit MODE signal selects Posit-8, Posit-16, or Posit-32 operation. Posit-8 mode activates four independent SIMD lanes, Posit-16 mode enables two paired lanes, and Posit-32 mode merges all lanes into a single wide datapath, enabling efficient handling of variable-length regime fields across precisions.

The complementor performs mode-aware two’s complement operations with no inter-lane carry propagation in Posit-8 mode, localized carry propagation in Posit-16 mode, and full-width carry propagation in Posit-32 mode. The SIMD LOD and logarithmic barrel shifter adapt dynamically to the selected precision for regime decoding and normalization. The mantissa multiplier employs a shared set of modified Booth multipliers, activating four partial products in Posit-8 mode, two parallel multiplier groups in Posit-16 mode, and full aggregation in Posit-32 mode, thereby avoiding datapath replication while preserving throughput.

\begin{figure}[!t]
    \centering
    \includegraphics[width=0.95\linewidth]{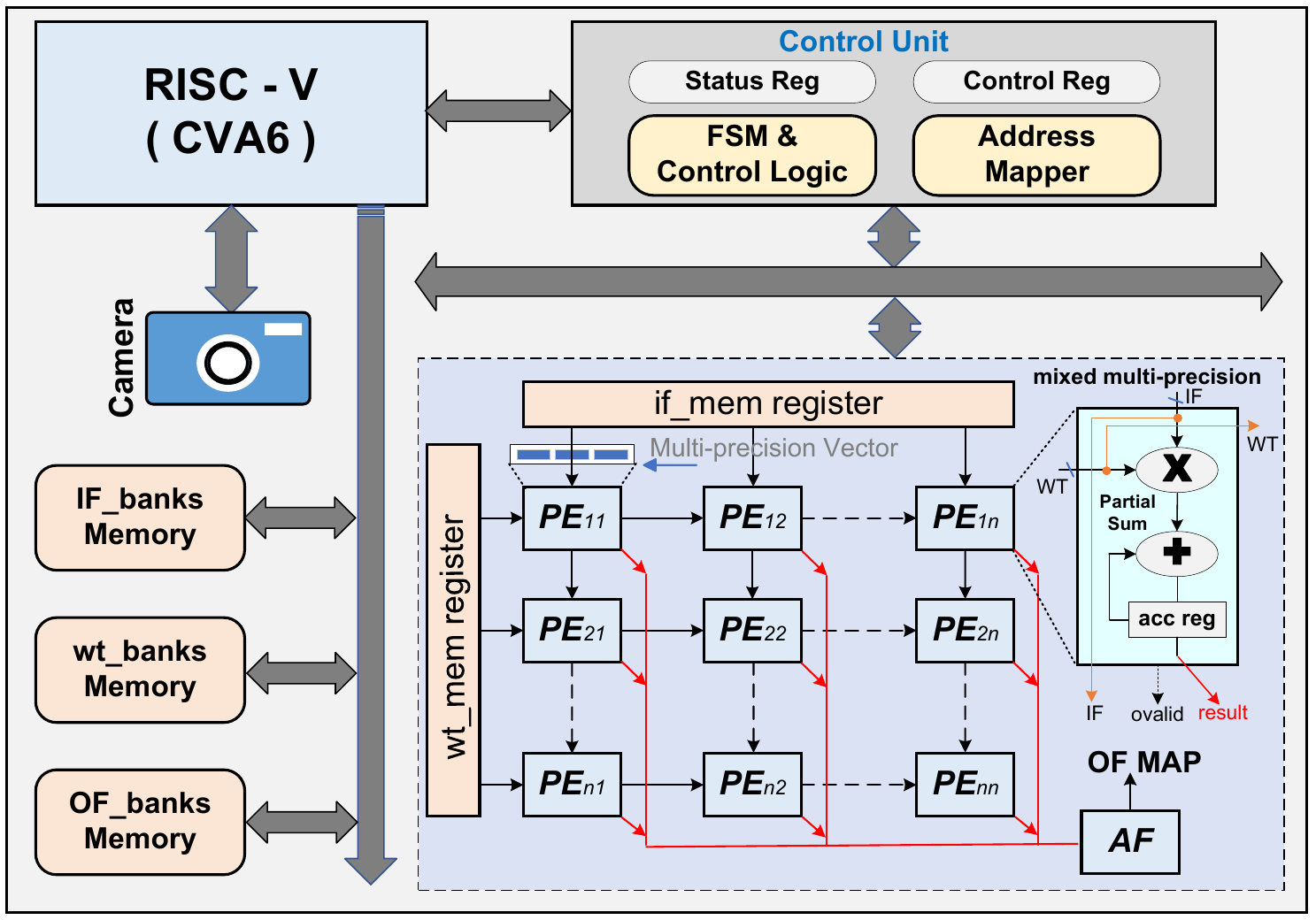}
    \caption{Detailed micro-architecture for SIMD Posit compute engine based systolic array architecture, Cheshire interface (CVA6)\cite{Cheshire}, control unit and memory banks.}
    \label{fig:rvsoc}
\end{figure}

Compared to a standalone Posit-32 MAC, the proposed SIMD Posit-8/16/32 architecture amortises hardware cost across precision modes by sharing critical datapath components, as shown in Fig. \ref{fig:comp_circ}. While introducing modest control and multiplexing overhead, SPADE achieves substantial gains in effective throughput, delivering up to 4× parallel MAC operations in Posit-8 mode and 2× throughput in Posit-16 mode without replicating full datapaths. Fig.~\ref{fig:comp_circ} highlights the key configurable building blocks of the SIMD Posit engine. At the system level, the processing element seamlessly integrates into a systolic array accelerator (Fig.~\ref{fig:rvsoc}), enabling scalable, energy-efficient mixed-precision DNN inference.

\section{Methodology and Evaluation}

This section outlines the methodology used to design, implement, and evaluate the proposed SIMD Posit MAC architecture. The evaluation spans RTL implementation and functional validation, FPGA deployment, ASIC synthesis across multiple technology nodes, hardware overhead analysis for multi-precision SIMD support, and inference accuracy evaluation on representative DNN workloads, collectively demonstrating the architectural efficiency, scalability, and numerical robustness of the proposed design.

All Posit MAC units were implemented in Verilog HDL and validated using the Xilinx Vivado 2018.3 Design Suite. The workflow includes RTL development, behavioral simulation, synthesis, timing analysis, and FPGA deployment on a Xilinx Virtex-7 device (xc7z010clg400). Functional validation was performed using 1000 randomized test cases spanning different sign combinations and dynamic ranges. Hardware outputs were cross-verified against the SoftPosit Python library for Posit(8,0), Posit(16,1), and Posit(32,2), with exact agreement observed in all cases, confirming correct Posit arithmetic implementation.

FPGA implementation results are summarized in Table~\ref{tab:fpga-comp-mac}. The proposed architecture demonstrates substantial resource savings over prior Posit MAC designs while eliminating DSP usage. The Posit(8,0) MAC achieves up to 45.13\% LUT reduction and 80\% slice reduction, while Posit(16,1) and Posit(32,2) achieve LUT reductions of 28.44\% and 17.47\%, respectively, along with significant register savings. The multi-precision SIMD MAC introduces only modest overhead compared to a standalone Posit(32,2) MAC, incurring a 6.9\% increase in LUTs and a 14.9\% increase in registers while enabling simultaneous multi-precision execution.

\begin{table}[!t]
\caption{FPGA Utilization, compared with State-of-the-art\\ SIMD MAC compute engines.}
\label{tab:fpga-comp-mac}
\renewcommand{\arraystretch}{1.25}
\resizebox{\columnwidth}{!}{%
\begin{tabular}{|l|l|l|l|l|l|}
\hline
Design & Precision & LUT & FF & Delay (ns) & Power (mW) \\ \hline
\multirow{4}{*}{This Work} & POSIT-8 & 366 & 41 & 1.22 & 93 \\ \cline{2-6} 
 & POSIT-16 & 1341 & 144 & 1.52 & 119 \\ \cline{2-6} 
 & POSIT-32 & 5097 & 544 & 2.45 & 402 \\ \cline{2-6} 
 & SIMD POSIT 8/16/32 & 5674 & 625 & 2.51 & 569 \\ \hline
ISCAS'25\cite{LPRE} & \begin{tabular}[c]{@{}l@{}}Approx. SIMD \\ Log Posit 8/16/32\end{tabular} & 4613 & 2078 & 6.2 & 276 \\ \hline
TCAS-II'24\cite{RPE} & SIMD INT4/FP8/16/32 & 8054 & 1718 & 4.62 & 296 \\ \hline
TVLSI'23\cite{MPE} & SIMD FP16/32/64 & 8065 & 1072 & 5.56 & 376 \\ \hline
TCAS-II'22\cite{UV-MAC} & POSIT-FP8/16/32 & 5972 & 1634 & 3.74 & 99 \\ \hline
\end{tabular}}
\end{table}

\begin{table}[!t]
\caption{ASIC Resources with CMOS 28nm, compared\\ with State-of-the-art SIMD MAC compute engines.}
\label{tab:asic-comp-mac}
\renewcommand{\arraystretch}{1.15}
\resizebox{\columnwidth}{!}{%
\begin{tabular}{|l|r|r|r|r|}
\hline
Design & \multicolumn{1}{l|}{Supply (V)} & \multicolumn{1}{l|}{Freq.(GHz)} & \multicolumn{1}{l|}{Area(mm\textsuperscript{2})} & \multicolumn{1}{l|}{Power (mW)} \\ \hline
This Work & 0.9 & 1.38 & 0.025 & 6.1 \\ \hline
TVLSI'25\cite{Flex-PE} & 0.9 & 1.36 & 0.049 & 7.3 \\ \hline
ISCAS'25\cite{LPRE} & 0.9 & 1.12 & 0.024 & 32.68 \\ \hline
TCAD'24\cite{DP-DAC} & 1.0 & 1.47 & 0.024 & 82.4 \\ \hline
TCAS-II'24\cite{FMA} & 1.0 & 1.56 & 0.022 & 72.3 \\ \hline
TCAS-II'24\cite{RPE} & 1.0 & 1.47 & 0.01 & 15.87 \\ \hline
TCAS-II'22\cite{UV-MAC} & 1.05 & 0.67 & 0.052 & 99 \\ \hline
\end{tabular}}
\end{table}

This additional hardware supports flexible execution modes of 1× Posit-32, 2× Posit-16, or 4× Posit-8 MAC operations per cycle, significantly increasing effective throughput for low-precision layers and improving the performance-area trade-off. To evaluate technology scalability, the proposed MAC was synthesized using Synopsys tools targeting TSMC 28 nm, 65 nm, and 180 nm nodes. At 28 nm, the MAC achieves 1.38 GHz at 0.9 V with 6.1 mW power consumption and an area of 2.548 µm\textsuperscript{2}. At 65 nm and 180 nm, the area scales to 11.54 µm\textsuperscript{2} and 90.95 µm\textsuperscript{2}, respectively, with proportionally low power and modest delay increases. Comparative ASIC results (Table~\ref{tab:asic-comp-mac}) and stage-wise analysis (Table~\ref{tab:stage-wise_comp}) confirm superior energy efficiency and area savings over state-of-the-art designs.

\begin{figure}[!t]
    \centering
    \includegraphics[width=0.95\columnwidth, height=52.5mm]{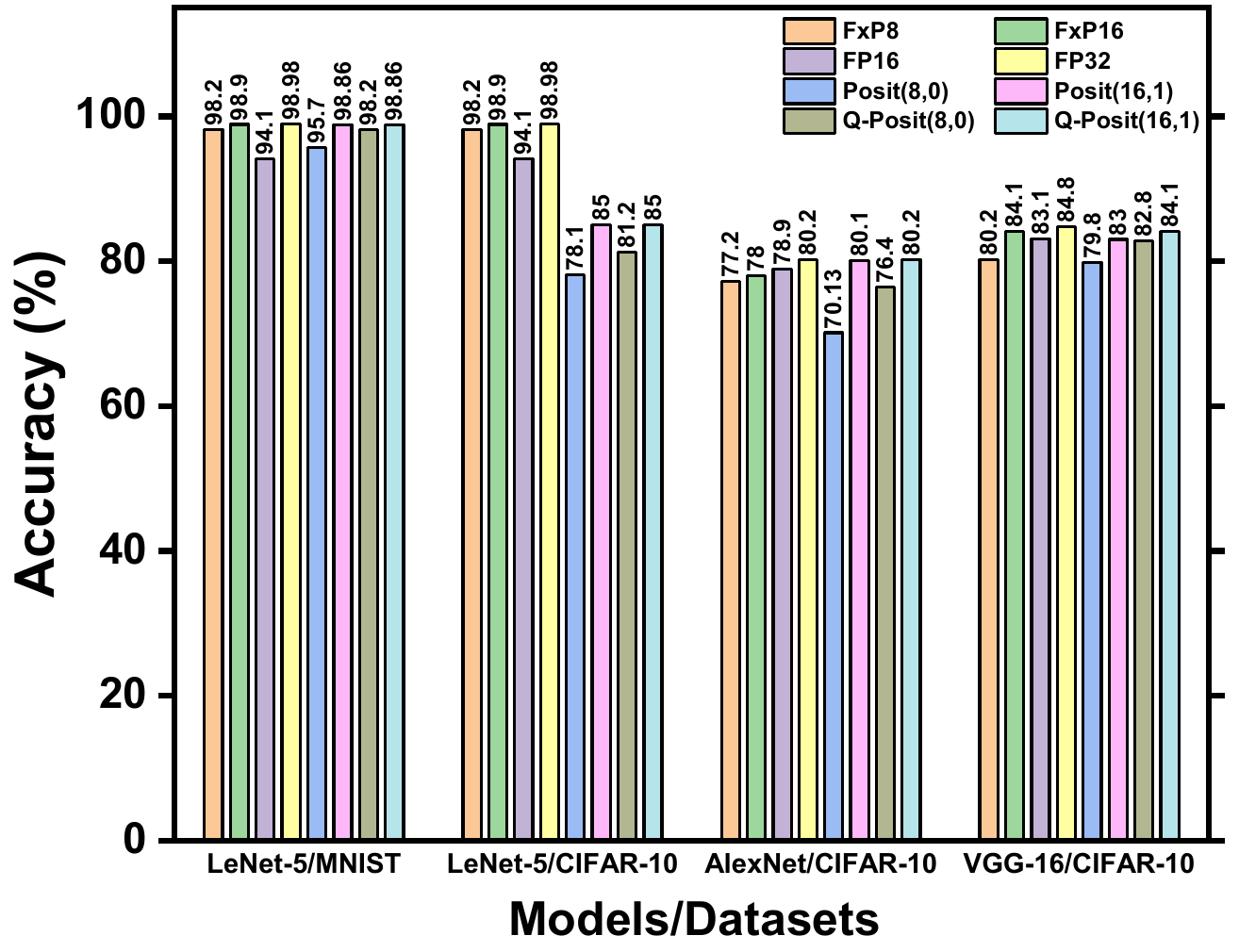}
    \caption{Comparative application accuracy for image classification, against prior works.}
    \label{fig:accuracy}
\end{figure}

Inference accuracy was evaluated on LeNet-5 (MNIST), a 5-layer CNN and AlexNet (CIFAR-10), VGG-16 (CIFAR-100), and a 4-layer CNN for alphabet recognition. As shown in Fig.~\ref{fig:accuracy}, SPADE maintains iso-accuracy relative to floating-point baselines while significantly reducing hardware cost, validating Posit arithmetic as a viable low-precision alternative for edge-AI workloads.

Overall, while standalone Posit MACs offer strong efficiency and numerical robustness, they lack adaptability across precision requirements. The proposed multi-precision SIMD Posit MAC overcomes this limitation by supporting Posit-8, Posit-16, and Posit-32 within a unified architecture. By combining effective throughput, operating frequency, and power consumption, a normalized throughput-per-watt metric shows that SPADE achieves up to 4× higher effective MACs/W in Posit-8 mode compared to standalone Posit-32 designs, with only 6.9\% LUT and 14.9\% register overhead.

\begin{table*}[!t]
\caption{Comparative evaluation for stage-wise resources with prior works.}
\label{tab:stage-wise_comp}
\renewcommand{\arraystretch}{1.30}
\resizebox{\textwidth}{!}{%
\begin{tabular}{|l|rr|rr|rr|rr|rr|}
\hline
\multirow{2}{*}{Design} & \multicolumn{2}{c|}{This Work} & \multicolumn{2}{c|}{TCAD'24\cite{DP-DAC}} & \multicolumn{2}{c|}{TCAS-II'24\cite{RPE}} & \multicolumn{2}{c|}{TVLSI'23\cite{MPE}} & \multicolumn{2}{c|}{TCAS-II'22\cite{UV-MAC}} \\ \cline{2-11} 
 & \multicolumn{1}{l|}{Area ($\mu$m\textsuperscript{2})} & Power (mW) & \multicolumn{1}{l|}{Area ($\mu$m\textsuperscript{2})} & \multicolumn{1}{l|}{Power (mW)} & \multicolumn{1}{l|}{Area ($\mu$m\textsuperscript{2})} & \multicolumn{1}{l|}{Power (mW)} & \multicolumn{1}{l|}{Area ($\mu$m\textsuperscript{2})} & \multicolumn{1}{l|}{Power (mW)} & \multicolumn{1}{l|}{Area ($\mu$m\textsuperscript{2})} & \multicolumn{1}{l|}{Power (mW)} \\ \hline
Input Proc. & \multicolumn{1}{r|}{3754} & 1.21 & \multicolumn{1}{r|}{\multirow{2}{*}{14735}} & \multirow{2}{*}{45} & \multicolumn{1}{r|}{\multirow{2}{*}{13432}} & \multirow{2}{*}{41} & \multicolumn{1}{r|}{\multirow{2}{*}{6575}} & \multirow{2}{*}{24.5} & \multicolumn{1}{r|}{8079} & 16.2 \\ \cline{1-3} \cline{10-11} 
\begin{tabular}[c]{@{}l@{}}Mantissa Mult. \&\\ Exponent Proc.\end{tabular} & \multicolumn{1}{r|}{10550} & 2.14 & \multicolumn{1}{r|}{} &  & \multicolumn{1}{r|}{} &  & \multicolumn{1}{r|}{} &  & \multicolumn{1}{r|}{22772} & 43.5 \\ \hline
Accumulation & \multicolumn{1}{r|}{5432} & 1.73 & \multicolumn{1}{r|}{3058} & 12 & \multicolumn{1}{r|}{5636} & 20 & \multicolumn{1}{r|}{1540} & 8.7 & \multicolumn{1}{r|}{13274} & 26 \\ \hline
Output Proc. & \multicolumn{1}{r|}{5120} & 1.03 & \multicolumn{1}{r|}{6320} & 25.5 & \multicolumn{1}{r|}{2849} & 11.4 & \multicolumn{1}{r|}{4914} & 26 & \multicolumn{1}{r|}{5855} & 26 \\ \hline
Total & \multicolumn{1}{r|}{24856} & 6.11 & \multicolumn{1}{r|}{24113} & 82.5 & \multicolumn{1}{r|}{21917} & 72.4 & \multicolumn{1}{r|}{13029} & 59.2 & \multicolumn{1}{r|}{49980} & 111.7 \\ \hline
\end{tabular}}
\end{table*}

\section{Conclusion \& Future Work}

This work presents a unified multi-precision SIMD Posit-based MAC architecture optimized for energy-efficient deep neural network acceleration. Standalone Posit MAC units for Posit (8,0), Posit (16,1), and Posit (32,2) demonstrate clear advantages over floating-point and fixed-point designs, achieving up to 45.13\% LUT reduction, 80\% slice reduction, and lower power consumption while maintaining full numerical accuracy validated against SoftPosit. To address the limited scalability of single-precision units, a multi-precision SIMD Posit MAC is developed that supports one 32-bit, two 16-bit, or four 8-bit Posit operations in parallel within a unified datapath. The architecture employs precision-configurable submodules, including the Complementor, LOD, Shifter, and Booth-based Mantissa Multiplier, enabling multi-precision execution with only 6.9\% additional LUTs and 14.9\% register overhead compared to a standalone Posit (32,2) MAC. Overall, the proposed SIMD Posit MAC delivers high throughput, compact area, and strong energy efficiency, making it well suited for next-generation edge-AI accelerators. Its regime-aware, precision-scalable design further provides a foundation for mixed-precision training and emerging transprecision workloads in edge and near-sensor AI systems.

\bibliographystyle{ieeetr}
\bibliography{bib}

\end{document}